 \documentclass[twocolumn, a4paper, 10pt, showpacs, prb, reprint, aps,superscriptaddress, amsmath, amssymb, amsfonts, floatfix]{revtex4-1}
\usepackage{amsmath, amssymb, latexsym}
\usepackage{amsfonts}
\usepackage{grffile}
\usepackage{graphicx}% Include figure files
\usepackage[utf8]{inputenc}
\usepackage{dcolumn}% Align table columns on decimal point
\usepackage{xcolor}

\DeclareMathOperator{\tr}{tr}
\DeclareMathOperator{\Tr}{Tr}

\begin{document}
	
	\title{Linearized spectral decimation in fractals}
\author{Askar A. Iliasov}
	\email{A.Iliasov@science.ru.nl}
	\affiliation{Institute for Molecules and Materials, Radboud University, Heyendaalseweg 135, 6525AJ Nijmegen, The Netherlands}
	
	\author{Mikhail I. Katsnelson}
	\affiliation{Institute for Molecules and Materials, Radboud University, Heyendaalseweg 135, 6525AJ Nijmegen, The Netherlands}
	
	\author{Shengjun Yuan}
	\email{s.yuan@whu.edu.cn}
	\affiliation{School of Physics and Technology, Wuhan University, Wuhan 430072, China}
	\affiliation{Institute for Molecules and Materials, Radboud University, Heyendaalseweg 135, 6525AJ Nijmegen, The Netherlands}
	
	%\date{\today}

\begin{abstract}
 In this article we study the energy level spectrum of fractals which have block-hierarchical structures. We develop a method to study the spectral properties in terms of linearization of spectral decimation procedure and verify it numerically. Our approach provides qualitative explanations for various spectral properties of self-similar graphs within the theory of dynamical systems, including power-law level-spacing distribution, smooth density of states and effective chaotic regime.  
\end{abstract}

	\maketitle
	
	%\tableofcontents
	
	\section{Introduction}
	
	Fractals were intensively studied in 80's of the last century. Recent developments with experimental techniques \cite{Polini_etal2013,
    Gibertini_etal2009, Shang_etal2015, Kempkes_etal2018}, open possibilities to study condensed matter systems with complex geometry structures like fractals on the atomic level. Theoretical and numerical works on fractals appeared recently include topics such as  conductivity and optical properties \cite{Veen2016,Veen2017, SongZhangLi2014, Westerhout2018}, localization \cite{SticAkh2016, KosKrzys2017}, topology of fractals \cite{Brzezetal2018, AgPaiShen2018, Pal_etal2019, PaiPrem2019,Bouzerar2020}, appearance of flat bands \cite{PalSaha2018, NandPal2015,  NandChak2016, Nandy2020}, and others \cite{Golmankhaneh2019, Akal2018}.

	One of the main features of a fractal is its hierarchical block structure, which repeats itself from one scale to another. It is known that for a simple fractal (such as Sierpinski gasket), the renormalization group on scale induces the spectral decimation procedure on the spectrum or the density of states, which can be interpreted as a direct renormalization on spectrum \cite{Shima1996}. However, it is not clear is there a general approach for a fractal with complex structure.

	A lot of fractal-like structures admit spectral decimation procedure, i.e., there is a connection between scale in real space and scale in the spectrum, and the whole spectrum is a limit set of iterations of some functions. But the properties of spectrum can be very different depending on the system. In some cases, the spectrum is a union of Cantor set with some degenerate eigenvalues \cite{Domany_1983}; in other cases, the limit spectrum can be a smooth function \cite{ShalmMor2019} like these in Dhar structures \cite{Dhar1977, Dhar1978}. For quasiperiodic potentials there were numerical investigations, which show that level-spacing-distribution has a power-law spectrum \cite{MachFuj1986, Nakaetal2005}. For iterations of non-linear functions it was shown that they also have power-law level-spacing distributions in some cases \cite{Nakaetal2005, IlKatSheng2019}.
	
    However, it is not clear how to determine the spectral behaviour of a fractal in general. Even if there exists a spectral decimation procedure, it is not enough to make a certain statement about spectrum behaviour.
	Of course, it looks almost impossible to build precise theory, but it is possible to build an effective theory neglecting details of a graph geometry.
	
	In this article we present linearized version of spectral decimation, which can be applied for block-built graphs. It is shown that with linearized spectral decimation functions, one can qualitatively describe level-spacing-distribution for hierarchical graphs and deduce possible phase transitions.
	One can hypothesize that the observed transitions can be interpreted as chaos-order transitions.
	
	The paper is organized as follows. In Sec. \ref{sec:tensorscale} we describe the algebraic motivation and the linearized version of the spectral decimation procedure, i.e., using dynamical system to generate the spectrum. In Sec. \ref{sec:dynsys} and \ref{sec:examples} we apply this dynamical system to some practical cases. In Sec. \ref{sec:dimprop} we discuss the connection of our approach to other physical properties such as the electronic conductivity. Finally a brief summary of our study is given in Sec. \ref{sec:sum}. Appendix \ref{sec:geometry} describes the geometrical interpretation of our model.
	
	\section{Tensor structure of scales}
	\label{sec:tensorscale}

	\subsection{Representation of fractals}
	
	A fractal can be described by one-particle tight-binding Hamiltonian:
	
	\begin{equation}	H= -\sum_{\langle ij\rangle} t_{ij} c^\dag_i c_j \,,
	\label{Eq:TBmodel}
	\end{equation}	
    which describes electrons with hopping between the nearest-neighbor $\langle ij\rangle$ sites of a fractal, $c^\dag_i$ and $c_j$ are creation and annihilation fermionic operators. We can regard this Hamiltonian as an adjacency matrix $A$ of a graph. The adjacency matrix is a square matrix $A$ such that its element $A_{ij}$ is one when there is an edge from vertex $i$ to vertex $j$, and zero when there is no edge (if an electron can jump from one site to another there is an edge connecting two sites).
	
	Let us consider a fractal with hierarchical block structure. This graph structure induces a block structure in adjacency matrix. For example, if $A_k$ is an adjacency matrix of $k$th iterations of a fractal, then the diagonal sub-matrices will be equal to $A_{k-1}$, which is an adjacency matrix of previous iteration. The non-diagonal sub-matrices represent connections between different blocks. If there is no connection between blocks on the first iteration there will be no connection further, and corresponding non-diagonal sub-matrices will be always zero. So, for $A_k$ we can write an expression using Kronecker product of matrices $\otimes$ (which has properties of tensor product):
	
	\begin{equation}
	A_k=A_{k-1}\otimes 1 + \sum_{\alpha} C_{k-1, \alpha}\otimes a_{\alpha}\,,
	\end{equation}	
	where matrices  $C_{k, \alpha}$ describe detailed connections between blocks, $a_{\alpha}$ are built from the adjacency matrix of the first iteration of a fractal $a=A_0$. Every matrix $a_{\alpha}$ has one non-zero component in the way that $a=\sum a_{\alpha}$. So, matrices $a_{\alpha}$ represent non-zero connections between different blocks of a fractal.
	
	To build matrices $C_{k, \alpha}$,we start from the second iteration:
	
	\begin{equation}\label{eq:A2}
	A_2=a\otimes 1 + \sum_{\alpha} c_{\alpha}\otimes a_{\alpha}\,,
	\end{equation}
	here matrices $c_{\alpha}$ define the detailed connections between different blocks of a fractal. Since fractals have self-similar structures, we can write that:
	
	\begin{equation}
	C_{k, \alpha}=\underbrace{c_{\alpha}\otimes c_{\alpha}...\otimes c_{\alpha}}_{k\,\, times}=c^{\otimes k}_{\alpha} \,.
	\end{equation}
	
    Then for the $k$th iteration of a fractal we have:
	
	 \begin{equation}\label{eq:fractal_construct}
	 A_k=A_{k-1}\otimes 1 + \sum_{\alpha} c^{\otimes (k-1)}_{\alpha}\otimes a_{\alpha}\,,
	 \end{equation}
	 
	 which can be also expressed as
	 
	 \begin{equation}\label{eq:fractal_construct_full}
	 A_k=a\otimes 1^{\otimes(k-1)} + \sum^{k-1}_{l=1}\sum_{\alpha} c^{\otimes l}_{\alpha}\otimes a_{\alpha}\otimes 1^{\otimes(k-l-1)}\,.
	 \end{equation}
	
	The above expression shows how adjacency matrices of fractals are constructed from basic blocks representing a rough hierarchical structure (via the first iteration of a fractal) and detailed connections between blocks. It is easy to check that fractals such as Sierpinski carpet and extended Sierpinski gasket can be constructed following equation \eqref{eq:fractal_construct_full}.
	
	\subsection{Spectral properties of tensor products}
	
	From algebraic point of view we can see that in some simple cases, different scales are decoupled. For example, in the case of Cartesian products $H\square G$ of graphs $H$ and $G$, its adjacency matrix is $A_{H\square G}=A_H\otimes 1 + 1\otimes A_G$, then the eigenvectors of this adjacency matrix are tensor products of eigenvectors of $A_H$ and $A_G$. Therefore, we obtain:
	
	\begin{equation}\label{eq:cartprodspec}
	A_{H\square G}(\psi_{H i}\otimes \psi_{G j})=(\lambda_{H i}+\lambda_{G j})\psi_{H i}\otimes \psi_{G j}
	\end{equation}	
	where $\lambda_H$ and $\lambda_G$ are eigenvalues of matrices $H$ and $G$. It indicates that the spectrum of Cartesian products of two graphs is the sum of each individual spectrum.
	
	A straightforward way to generalize the result of equation \eqref{eq:cartprodspec} is to increase the number of summands and the number of tensor products as the following:
	
	\begin{equation}
	A_{h}=\sum_{\alpha}\Pi^{\otimes}_{\beta}h_{\alpha\beta}\,,
	\end{equation}	
	where ${h_{\alpha\beta}}$ is a set of matrices. If $[h_{\alpha\beta},h_{\alpha'\beta}]=0$ for every fixed $\beta$, i.e., all matrices of the same scale commute. Then eigenvectors of the matrix $A_{h}$ are tensor products of eigenvectors of $h_{\alpha\beta}$. The spectrum will be sums of products of the corresponding eigenvalues.
	
	Unfortunately, this approach can not be applied directly to fractals, because of the non-commutativity of matrices $c_{\alpha}$ and $a_{\alpha}$ in equation (\ref{eq:fractal_construct_full}). But one can make an estimation by a kind of algebraic averaging, which can be interpreted as a mean-filed theory, to overcome the difficulty raised by the noncommutative matrices.
	
	In order to do this, let us first consider a matrix with the form:
	
	\begin{equation}\label{eq:fractal_sum}
	 A^{sum}_{k}=a\otimes 1^{\otimes(k-1)}+\sum^{k-1}_{l=1}c^{\otimes l} \otimes a \otimes 1^{\otimes(k-l-1)}\,,
	\end{equation}
	
	where $c=\sum c_{\alpha}$ and $a=\sum a_{\alpha}$. The matrix $A^{sum}_{k}$ has a similar structure as $A_{k}$ in Eq. \eqref{eq:fractal_construct_full} with $a_{\alpha}$ and $c_{\alpha}$ replaced by their sums. One can also see, that $A^{sum}_{k}$ can be written as a sum of various graphs with structure of $A_k$, but with all possible permutations of indexes $\alpha$. If we assume that different variants of organizing connections between blocks are equal, i.e., we neglect detailed geometry, then we can write $c_{\alpha}=\epsilon c$, where $\epsilon^{-1}=n_{c}$ is the number of $c_{\alpha}$. Having done that, we obtain a weighted version of Eq. (\ref{eq:fractal_sum}):
	
	\begin{equation}\label{eq:fracta_sum_weighted}
	\hat{A_{k}}=a\otimes 1^{\otimes(k-1)}+\sum^{k-1}_{l=1}\epsilon^l c^{\otimes l} \otimes a \otimes 1^{\otimes(k-l-1)}
	\end{equation}
	
	Now, the only condition remaining that needed to be satisfied is the commutativity of $a$ and $c$. However, in some cases such as extended Sierpinski gasket, $c$ is proportional or even equal to $a$, then we can calculate the spectrum analytically as the following.
	
	If $c=a$, the spectrum of $\hat{A}_k$ is given by the formula:
	
	\begin{gather}\label{eq:sum_spectrum}
	\sigma(\hat{A}_k)=\{\lambda_{i_1}+\epsilon\lambda_{i_1}\lambda_{i_2}+\epsilon^2\lambda_{i_1}\lambda_{i_2}\lambda_{i_3}\\+\ldots+\epsilon^{k-1}\lambda_{i_1}\lambda_{i_2}\cdot\ldots\cdot\lambda_{i_k}\}\nonumber
	\end{gather}
	
	where $\lambda_{i}$ are eigenvalues of the matrix $a$. Lower indices mean that to obtain one specific eigenvalue in $\sigma(\hat{A}_k)$, one needs to choose the $k$ eigenvalues of $a$ $\{\lambda_{i_1}, \lambda_{i_2},\ldots,\lambda_{i_k}\}$ and substitute them into the expression of Eq. (\ref{eq:sum_spectrum}). All possible choices give the whole spectrum of the $\hat{A}_k$.
	
	One can notice a similarity between Eq. \eqref{eq:sum_spectrum} and the conventional renormalization approach in quantum field theory \cite{Collins1984}. Because of the non-commutativity of block matrices $c_{\alpha}$, the correct spectral decimation functions are non-linear. We approximate a non-linear function by a linear one as an analogy to the one-loop approximation, and then iterate this linear function repeatedly in order to include all scales.

	\section{Corresponding dynamical system}
	\label{sec:dynsys}
	
	In this way we can consider a qualitative approximation to the spectral renormalization group, which is also called spectral decimation. This approximation is a dynamical system obtained as a multivalued linear function, with slopes equal to eigenvalues of a simple block ${\lambda_i}$ normalized to the number of connections between blocks.
	
	Although it is not possible to represent Eq. (\ref{eq:sum_spectrum}) as a dynamical system without additional normalization. Nevertheless, our approach is simple and does not influence on properties of the spectrum.
	Non-normalized eigenvalues $x_{k+1}$ on $k+1$th iteration are obtained from eigenvalues on $k$th iteration by action of functions $F_{i}$:
	
	\begin{equation}\label{eq:model_DS}
	x_{k+1}=F_{i}(x_k)=1+\epsilon \lambda_{i}x_k
	\end{equation}
	
	The resulting spectrum can be obtained via formula $\sigma(\hat{A}_k)=\{\lambda_{i}x_{k}\}$ with $x_{0}=1$. Another way to express the spectrum $\sigma(\hat{A}_k)$ is to consider $k+1$ iterations of (\ref{eq:model_DS}) in the following:
	
	\begin{equation}\label{eq:spectrum_DS}
	\sigma(\hat{A}_k)=\{\frac{(F_{i_{k+1}}\circ F_{i_{k}} \circ \ldots \circ F_{i_{1}})(1) -1}{\epsilon}\}
	\end{equation}
	
	The Eq. \eqref{eq:spectrum_DS} shows that the spectrum is obtained from the limit set of translation and rescaling. Statistical properties of the spectrum are independent of them, and in order to study spectral properties we can consider just the dynamical system without the last step of translation and rescaling.
	
	One can regard the spectrum as the splitting process of eigenvalues on each iteration with a weight factor $\epsilon$. The splitting can be represented as a tree, starting from the eigenvalues of the matrix $a$ of the building block, and in each iteration every eigenvalue splits to number of points with numbers equal to the rank of the matrix $a$. Despite simplicity of the process the model already demonstrates non-trivial structure of the spectrum.
	
	The level-spacing distribution can be calculated straightforward, when there is no intersections between branches in the tree. This condition depends also on properties of eigenvalues. If the case, when all eigenvalues of the matrix $a$ are positive, the condition of absence of the intersections between branches is: 
	
	\begin{equation}\label{eq:stability_cond}
	\min|\lambda_i-\lambda_{i+1}|<\frac{1}{1-\epsilon\lambda_{\max}}-\frac{1}{1-\epsilon\lambda_{\min}}
	\end{equation}
	
	If there is no intersection of branches, the level-spacing distribution $P(s)$ follows power-law distribution, becomes $\infty$ at $s=0$. Precisely, $P(s)$ is a bunch of delta-functions with power-law envelope.
	
	For the cases when number of intersections is small, if we increase the weighting factor $\epsilon$, these delta-functions begin to smear and drift closer to each other and the slope of the level-spacing distribution $P(s)$ increases. Then at some critical point, when smeared delta-functions are close enough to each other, there is a transition to another profile which looks more like the statistics of disordered systems.
	
	\begin{figure}
		%\centering
		\includegraphics[width=0.4\textwidth]{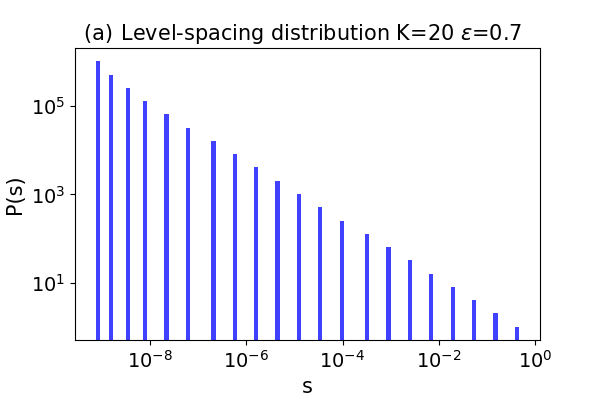}
		\includegraphics[width=0.4\textwidth]{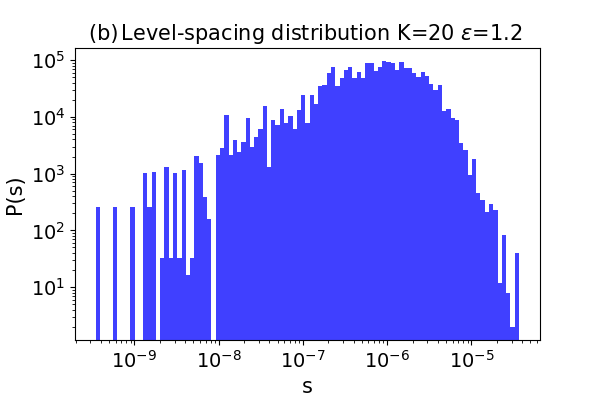}
		\caption{Level-spacing distribution $P(s)$ for a simple model of dynamical system with two eigenvalues $\{-1/2, 1/2\}$ after 20 iterations, exact power-law and after transition point. $\epsilon=0.7$ in a) and $\epsilon=1.2$ in b)}
		\label{fig:LSD_test_power-law}
	\end{figure}

	In order to show that kind of transition, let us consider first the simple model with two eigenvalues $\{-1/2, 1/2\}$. If $\epsilon<1$, one can obtain a power-law level-spacing distribution (see Fig. \ref{fig:LSD_test_power-law}a):
	
	\begin{equation}\label{eq:toy_mod_power-law}
	P(s)\sim (\frac{s}{1-\epsilon})^{\frac{\ln2}{\ln \epsilon  - \ln 2}}
	\end{equation}
	
	At the critical point $\epsilon=1$, the limit set continuous in the interval $[-1,1]$. The limit set becomes continuous, because On each iteration, all eigenvalues are equidistant from the neighboring eigenvalues and therefore a power-law distribution becomes a delta-function, which drifts to zero with increasing the number of iterations.
	
	If $\epsilon>1$, then there is no singularity in $P(s)$ at $s=0$, because of the mixing of tree branches (see an example shown Fig. \ref{fig:LSD_test_power-law}b). The exact power-law symmetry of consequent splitting is broken.
	
	Despite the simplicity of above model, the features of general non-linear iterations of functions should be captured correctly. It can be understood in the following. If we consider the invariant interval of a dynamical system, then there are two possibilities: invariant interval contains a gap or not. If there is a gap, then after one iterations this gap will be mapped into another one of smaller size and so on. Then the limit set is a Kantor set, and in many cases it has a power-law level-spacing distribution \cite{IlKatSheng2019}. This case corresponds to the non-intersection of branches in linearized version. If there is no gap, then the distribution of points after one iteration become can be effectively more chaotic, which corresponds to intersection of different branches. Thus we can distinguish three regimes of the dynamical system: fractal (without branches intersection), qualitatively chaotic (with intersections) and the one corresponding to the transition point between these two. One can speculate that a system with smooth profile of density of states corresponds exactly to the transition point between intersection and non-intersection regimes.

	\section{Examples}
	\label{sec:examples}
	
	\begin{figure}
		\centering
		\includegraphics[width=0.25\textwidth]{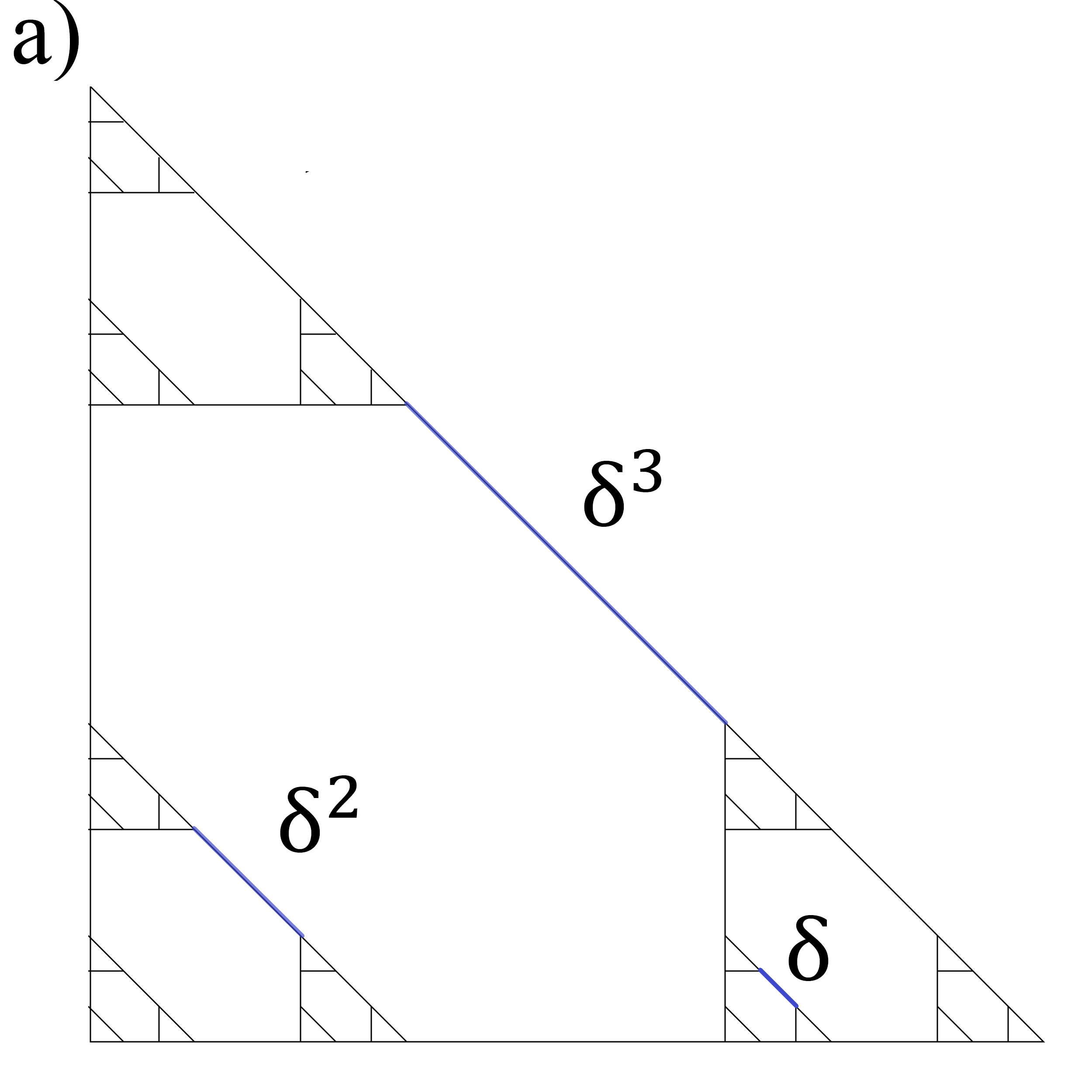}
		\includegraphics[width=0.25\textwidth]{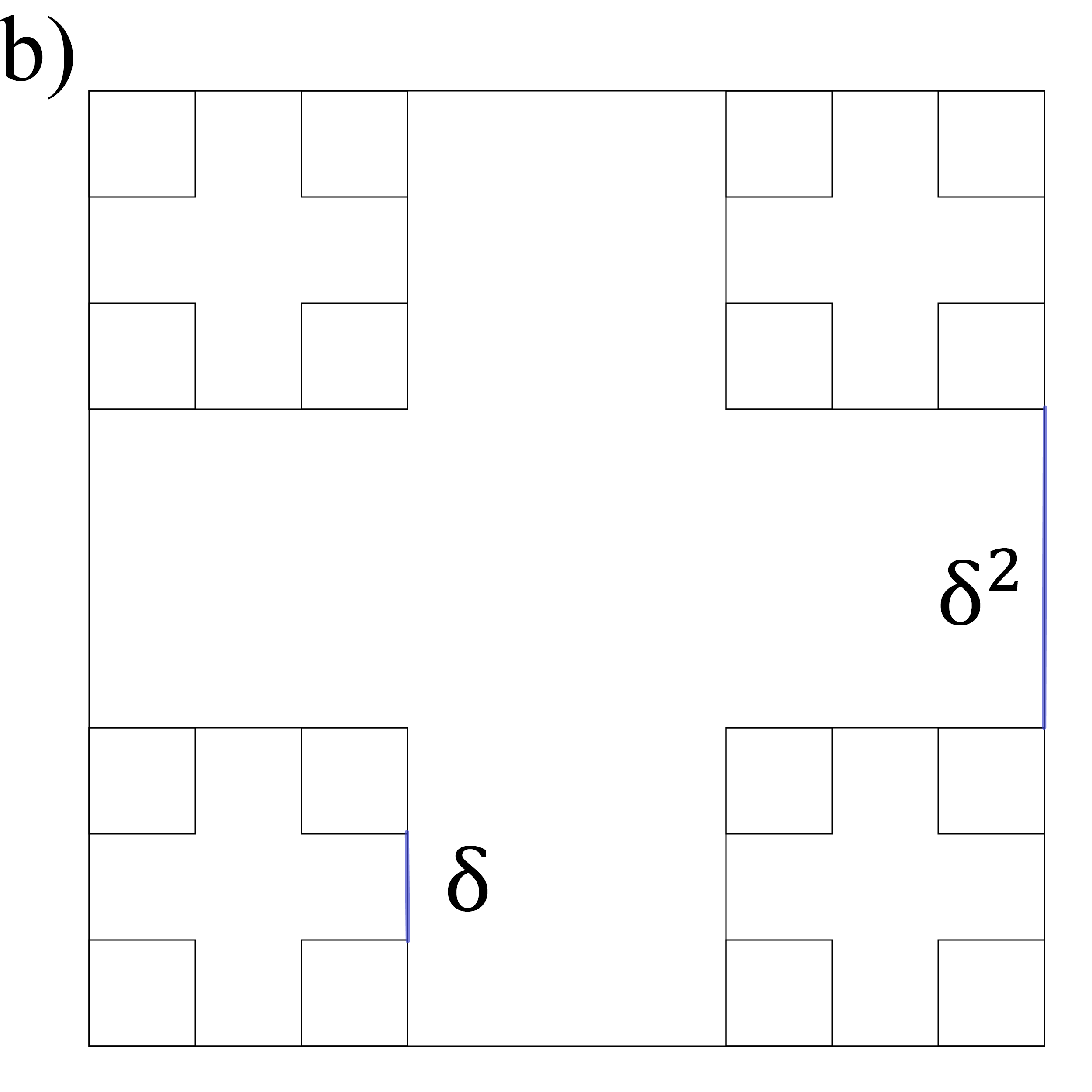}
		\caption{3 iterations of extended Sierpinski carpet and 2 iterations of extended square. The blue edges correspond to the neighboring weight. All edges of the same scale have the same weight.}
		\label{fig:Sierp_mod}
	\end{figure}

   	%\begin{figure}
   	%\centering
   	%\includegraphics[width=0.5\textwidth]{Sierpcarpet_4_1edges_eps=0.3_k=6.png}
   	%\includegraphics[width=0.5\textwidth]{Sierpcarpet_4_1edgesDS_eps=0.3_k=6.png}
   %	\caption{Spectrum of 6 iterations of the hierarchical graph with square blocks and of 6 iterations of dynamical system}
   	%\label{fig:Sierp_4_1_spectr}
   %\end{figure}
   
   	\begin{figure}
		%\centering
		\includegraphics[width=0.4\textwidth]{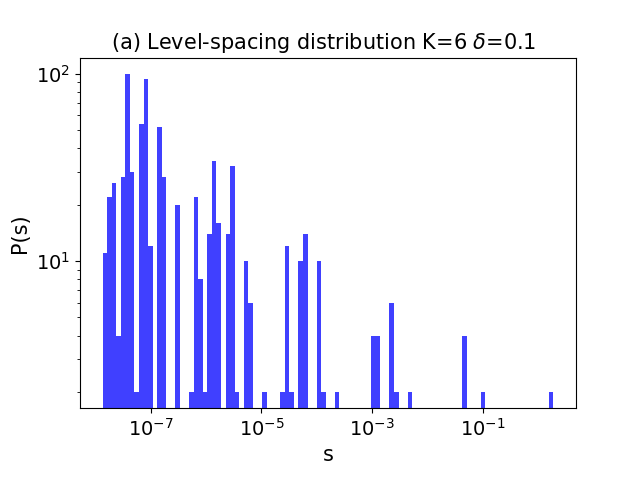}
			\includegraphics[width=0.4\textwidth]{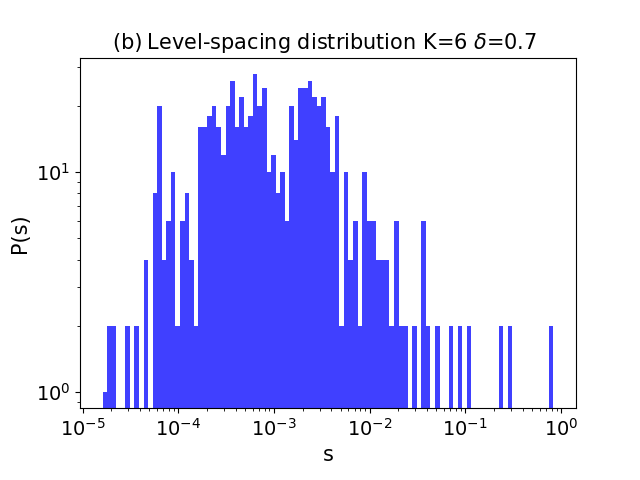}
		\caption{Level-spacing distribution of 6 iterations of the hierarchical graph with square block, $\delta=0.1$ and $\delta=0.7$. The top demonstrates power-law behaviour, the bottom one does not have obvious power-law dependence. }
		\label{fig:LSD_Sierp_4_1_delta}
	\end{figure}

	As was mentioned before, our approximation such as this presented in Eq. \eqref{eq:sum_spectrum} works better if connections between blocks are less. One of the best examples is to use cycles with one connection edge for a neighbor and add scaling parameter $\delta$ between blocks like in our previous studies in Ref. \onlinecite{IlKatSheng2019} using extended Sierpinski gasket shown in Fig. \ref{fig:Sierp_mod}, here one cycle consists of 3 vertices, therefore $\epsilon=\delta/3$. Another example of fractal that we studied is extended square with one connection between two blocks, therefore $\epsilon=\delta/4$.
	
	For a fractal of extended square the building block is a square with four sites described by Hamiltonian Eq. \eqref{Eq:TBmodel} with $t=1$. The eigenvalues of this building block are $\{-2, 0 , 2\}$, therefore, the dynamical system is very similar to the simplest case considered in the previous section. The transition point is equal to $\epsilon=0.25$, which corresponds to $\delta=1$ with critical exponent $2\ln2/\ln(2/9)$ (which is $\simeq-0.92$). This is the case without additional weighting on the edges. The critical point in the case of extended Sierpinski gasket is $\epsilon=1/3$, which also corresponds to $\delta=1$. This can be shown by an analysis of invariant interval of the dynamical system.
	
	We can see a power-law distribution with smeared peaks for the extended Sierpinski gasket in Fig. \ref{fig:LSD_Sierp_4_1_delta}a (top picture) on the contrary In Fig. \ref{fig:LSD_Sierp_4_1_delta}b, there is no power-law dependence. If there is no intersections in the splitting tree, the power-law spectrum is exact even with finite number of iterations. However, if there are intersections, delta-functions in $P(s)$ are smeared (as in the case of exact spectrum), therefore if $s$ is close to $0$, the level-spacing distribution function $P(s)$ will be determined by tails of smeared delta-functions. Thus, a power-law dependence appears only in some range, even before $\delta$ approaches transition point.
	
	      	\begin{figure}
   	%\centering
   	\includegraphics[width=0.4\textwidth]{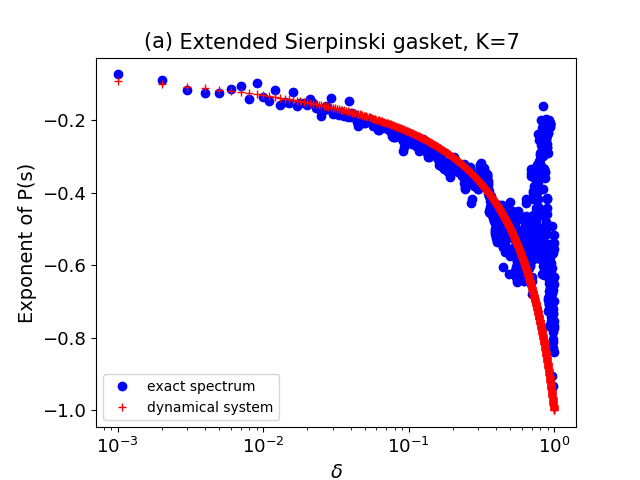}
   	\includegraphics[width=0.4\textwidth]{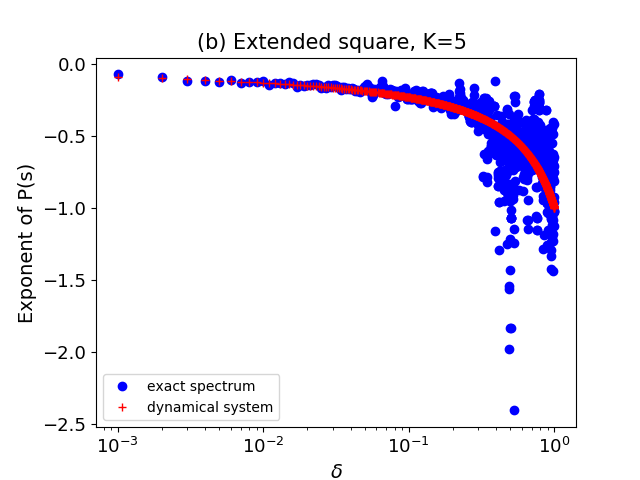}
   	\caption{Dependence of the exponent of power-law distribution as a function of $\delta$ for extended Sierpinski gasket (with $7$ iterations) and extended square (with $5$ iterations). The blue dots are obtained by exact diagonalization and red cross are obtained by our approaches with dynamical system}
   	\label{fig:splitting_dependance}
   \end{figure}
	
	The above explanation is verified by our numerical calculations. In Fig. \ref{fig:splitting_dependance}, the exponents of power-law level-spacing distribution of fractals and their corresponding dynamical systems are shown. We used $7$ iterations of extended Sierpinski gasket and $5$ iterations of extended square, and the same number of iterations for the dynamical systems. We calculated the exponent of power-law for different $\delta$ using linear regression in log-scale before the level-spacing distribution reaches the maximum i.e. we made a cut off on small $\delta$.  The dots of exact spectrum for small $\delta$ demonstrate clear power-law behaviour, and we see that with increased values of $\delta$ there are large fluctuations in the exponents.
	
\begin{figure}
   	%\centering
   	\includegraphics[width=0.4\textwidth]{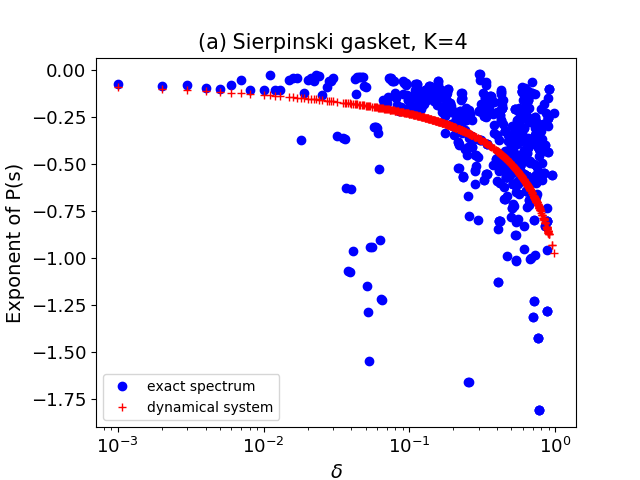}
   	\includegraphics[width=0.4\textwidth]{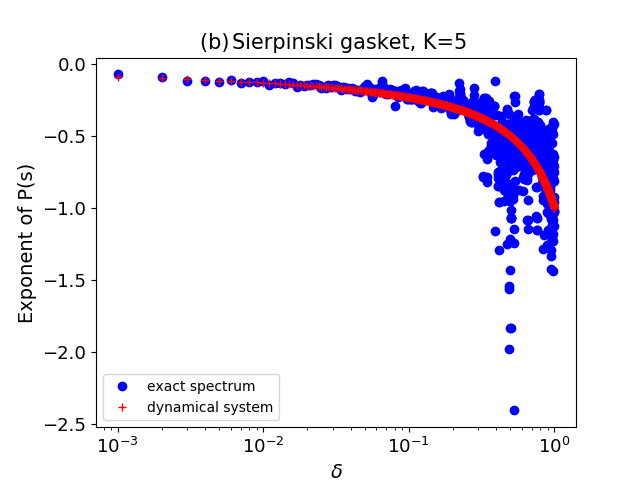}
      \includegraphics[width=0.4\textwidth]{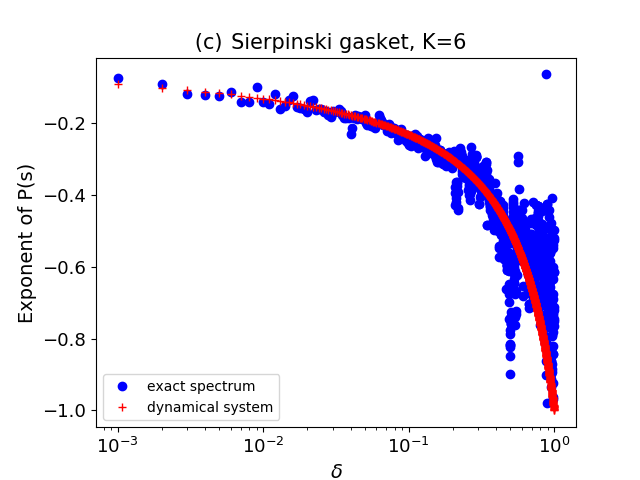}
            \includegraphics[width=0.4\textwidth]{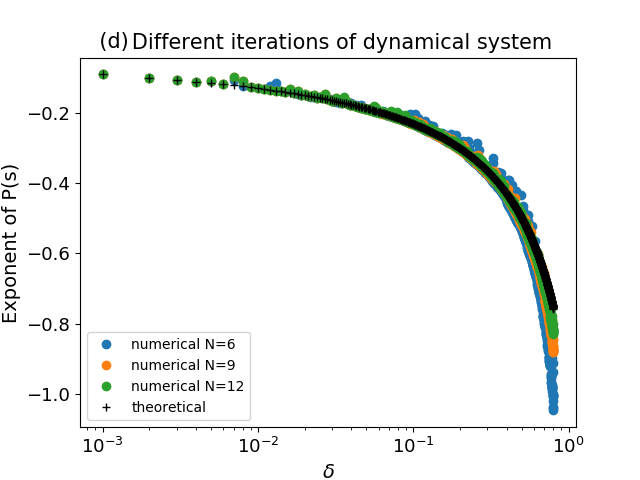}
   	\caption{(a-c) The comparison between exact spectrum and results obtained for the dynamical system with different iterations in extended Sierpinski gasket. d) Comparison between  numerical and theoretical results for the dynamical system. }
   	\label{fig:num-test}
   \end{figure}

	When we compare the results obtained from dynamical system to the exact spectrum, they match well for small values of $\delta$, in some cases even for $\delta>0.1$. In Fig. \ref{fig:num-test} we show more results for different iterations of extended Sierpinski gasket and corresponding dynamical system. One can see that with increasing the number of iterations the agreement between two approaches is also increased. Therefore we conclude that despite the fact that there should be an exact power-law for a full fractal, and if we consider only finite iterations, this power-law is not evident.
	
	The large fluctuations in the exact spectrum when $\delta$ approaches its critical point can be understood in the following. Around the critical point the effective splitting converges very slow with number of iterations, so the number of iterations should be very large. This issue is demonstrated in Fig. \ref{fig:num-test} d), in which we compare numerical results for various iterations of the dynamical system described by Eq. \eqref{eq:toy_mod_power-law} and its theoretical predictions. We see that the numerical method with finite iterations always gives larger exponent than theoretical prediction in the vicinity of critical point. However, the accuracy of numerical calculation increases with increasing of the number of iterations.

	\section{Dimension properties}
	\label{sec:dimprop}
	
    In this section we study dimension properties of a graph and its spectrum, and their connections to level-spacing statistics. We also partially explain the results obtained in Ref. \onlinecite{Veen2016}, where the authors found connection between the dimension of conductivity spectrum and the geometry dimension of Sierpinski fractals.
    
    In this section we use the notion of Hausdorff dimension $d_{H}$. This dimension is a generalization of a topological dimension for non-regular geometric sets such as fractals. The full mathematical definition is rather complicated, however, in many cases, Hausdorff dimension admits a simple description.
    Suppose that $N(\epsilon)$ is the number of boxes of side length $\epsilon$ required to cover the set. Then the Hausdorff dimension can be calculated as:
    
    \begin{equation}
     d_{H}=\lim _{\epsilon \to 0}{\frac{\log N(\epsilon )}{\log(1/\epsilon )}}
    \end{equation}
     
    Roughly speaking, the dimension is the exponent relating the volume of a set with its characteristic linear size $V\sim L^{d_{H}}$, which is what one would expect in the case of a smooth space.
	
	\subsection{Sample and spectrum}

	First, let's discuss briefly Hausdorff dimension of a spectrum. For a power-law spectrum, there are obviously gaps on all possible scales, therefore Hausdorff dimension can not be equal to one. Actually, one can extend the idea of gaps in all scales as a criteria of fractional dimension. However, there is a subtlety in the limit procedure.
	
	Let's consider the same toy model with two eigenvalues $\{-1/2, 1/2\}$, as the one we have studied. We have seen that there are two regimes with different properties depending on the value of $\epsilon$: $\epsilon<1$ and $\epsilon>1$. The spectrum for $\epsilon<1$ is Cantor set and Hausdorff dimension is $d_s =-\ln 2/(\ln\epsilon-\ln2)$. One may notice that Hausdorff dimension of the spectrum corresponds to the exponent in the power-law of level-spacing distribution shown in Eq. \eqref{eq:toy_mod_power-law}, i.e. $P(s)\sim s^{-d_s}$. %(the result obtained here is different from the work \cite{Nakaetal2005} by absence of $-1$)%
	 One can assume that the same result should hold for multiscale Cantor set (Cantor set, which is obtained by deleting intervals of various fractions), which also corresponds to the approximation for a general spectral decimation function \cite{IlKatSheng2019}.
	We want to remind that Hausdorff dimension discussed here is not the spectral dimension of the density of states. 
	
	The Hausdorff dimension of the spectrum of above simple model can be obtained from the relation $d_s=\ln(2)/\ln((1-\Delta_1/\Delta)/2)$, where $\Delta=1/(1-\epsilon/2)$ is the energy range of the spectrum (or invariant set of dynamical system) and $\Delta_1=(1-\epsilon)/(1-\epsilon/2)$ is the largest gap in the limit set of dynamical system. This formula can be understood in the following way. Since our dynamical system is linear, after the second iteration the biggest gap $\Delta_1$ repeats itself on the lower scale with some constant scaling factor. These new gaps repeat again with the same scaling factor and so on. Thus we can deduce the Hausdorff dimension from the first gap alone.
	
	If $\epsilon=1$, as was discussed before, in the limit set when the number of iterations approaches infinity, the values spread completely over $[-1,1]$, and therefore the Hausdorff dimension is $1$. If $\epsilon>1$, gaps become smaller and smaller after each iterations (one can see this from Fig. \ref{fig:LSD_test_power-law}), the limit set doesn't contain any gaps, and its Hausdorff dimension is also $1$. %This also can explain the result of the article \cite{Hernando2015}, where the level-spacing distribution of Sierpinski carpet was investigated. Authors obtained that $P(s)\sim s^{-1}$ for Sierpinski carpet. So, we can conclude that Sierpinski carpet corresponds to the transition case between power-law distribution and chaotic regime.
	
	The estimation of Hausdorff dimension of a hierarchical graph is more difficult, since correct value is related to embedding of a graph into a plane. However, it is possible for fractals with building blocks, which can tessellate an n-dimensional space, i.e. for a 2-dimensional plane they are triangle, square and honeycomb lattices. Furthermore there could be another problem occurs, when one try to include number of connections between blocks into account. Nevertheless, we can estimate the Hausdorff dimension by the following procedure. Basically, the concept behind the dimension is how many new copies appear, when we increase a length of a sample, $N_{new}\sim l^{d_G}$, where $d_G$ is a dimension of the sample. Therefore, one of key issue is an estimation of a proper choice of a length change. If there is an embedding into a space with integer dimension, it can be obvious. In general we need to work only with number of vertices $n_v$ and number of connections $n_c$. The number of connections is related to the effective length and number of vertices determines number of new copies $N_{new}$. Hence, we can estimate the dimension of a graph $\Gamma$ as $d_\Gamma\sim\ln n_v/ \ln 2 n_c$. For Sierpinski gasket we have $n_v=3$ and $n_c=1$, and we obtain $d_{\Gamma}\sim \ln3/\ln2$, which is the correct result.
	
	For hierarchically weighted graphs, one can consider additional weighting $\delta$ as in the previous section and obtain effective dimension  $d_\Gamma\sim\ln n_v/ \ln (2 n_c/\delta$). Furthermore, one can relate the dimension of the spectrum in the previous section and estimated dimension of a weighted fractal square. For this system we have $n_v=4$ and $n_c=1$, and we obtain $d_s=\ln 2/(\ln2-\ln \delta)$ and $d_\Gamma\sim2\ln 2/ (\ln2 -\ln\delta)$. $d_s$ and $d_{\Gamma}$ are not the same, but they differ only on some multiplier constant. However, we see that there is a deep relation between the dimension of spectrum and the dimension of a fractal. For example, we can notice that if $\delta\to0$ then both dimensions $d_{\Gamma}$ and $d_s$ goes to zero. Therefore we can conclude that systems with small Hausdorff dimension should have power-law level-spacing distribution.

	\subsection{Conductance}
	
    The conductance of a fractal can be calculated via Landauer formula \cite{Datta1997}.
	
	\begin{equation}
	G(E)_{ll'}=\frac{e^2}{h}\Tr(\Gamma_l G^{r}_S \Gamma_{l'}G^{a}_S)
	\end{equation}	
    where $l$ and $l'$ are indices correpsonding to leads, $G^{r}_S$ and $G^{a}_S$ are retarded and advanced Green functions,  $\Gamma_l$ and $\Gamma_{l'}$ take into account corrections to the self-energy regarding interaction with leads. Green functions have poles at points in the spectrum.
	
	Dimension of conductance as a function of energy $G(E)$ (i.e. dimension of the graph of $G(E)$) is related to the dimension of the spectrum of a sample. If there is no correlation between eigenstates, the dimension of $G(E)$ equals to the dimension of discontinuity points of $dG/dE$ (which is equal to the dimension of the spectrum) plus one. The correlations between eigenstates will smooth discontinuities.
	
	Therefore, we arrive at:
	
	\begin{equation}\label{E:conddim_ineq}
	d_{H}(G)\leq 1+d_{s}
	\end{equation}
	
	In our approach based on dynamical system, all eigenvectors are just tensor products of eigenvectors of a building block. Thus, all scalar products of eigenfunctions and matrices of leads $\Gamma_l$, $\Gamma_l'$ can be calculated and the conductivity will have non-regular fractal structure on all scales. At every pole of Green functions there is a discontinuity and the Eq. \eqref{E:conddim_ineq} becomes an equality within the considered approximation. As we discussed in previous section, that if a sample has a power-law distribution of $P(s)$, then its geometry dimension can be expressed as the dimension of spectrum with some multiplier (see the case of weighted fractal square). This multiplier depends on structure of building block (its' eigenvalues) and therefore, there is no universal formula between $d_{H}(G)$ and $d_\Gamma$.
	
	A subtle case appears at the transition point, when $d_s$ is close to $1$. In this case, inequality expressed in Eq. \eqref{E:conddim_ineq} is trivial and the dimension of spectrum provides no information about the conductance. Effectively, the spectrum is dense and Green functions have singularities on a continuous interval. The studied Sierpinski carpet in Ref. \onlinecite{Veen2016} seems to be this case.

	\section{Summary and Discussion}
	\label{sec:sum}
	
	In this work we considered the linearised version of spectral decimation within an approach based on the dynamical system for hierarchical graphs with block structures. We demonstrated that the power-law level-spacing statistics appeared in some fractals is closely related to their geometry. Our approach to calculate the level-spacing distribution shows different behaviour depending on the fractal structure. It was shown that the level-spacing distribution can have strictly a power-law behaviour or resemble behaviour of a quantum chaotic system.
	
	 The power-law spectrum is closely connected to ramification number of a fractal, however, the actual distinction is quite subtle. There could be infinitely ramified fractals with power-law spectra, as well as finitely ramified with spectra closer to disordered systems. The correct analysis of possible statistical properties should require an individual consideration in each case, since it depends on the eigenvalues of the building block of an hierarchical structure.
	 
	 Our approach based on the dynamical system can also explain the results concerning topological effects in fractals. It is well known that there is quantum Hall effect in 2D and it disappears in 1D. In Refs. \onlinecite{Brzezetal2018, IlKatSheng2020, Fremlingetal2020}, it was shown that Chern numbers as well as Hall conductivity become partially quantized in non-integer dimensions. In view of the present work, the actual transition from quantized topological properties in 2D to their destruction in 1D could be followed from the changes of hierarchical block structure of a graph and their corresponding dynamical system on its spectrum.
	
	To sum up, from the perspective of considered estimation, we can regard random graphs as deformations of graphs with block hierarchical structure. The additions of various building blocks and variations of connections between them lead to mixing of splitting branches, and therefore the power-law statistics disappears and the system becomes closer to a disordered system.
	
	\section*{Acknowledgements}
	
	We are thankful to Andrey Bagrov for the helpful discussions.
	This work was supported by the National Science Foundation
	of China under Grant No. 11774269 and by the Dutch
	Science Foundation NWO/FOM under Grant No. 16PR1024
	(S.Y.), and by the by the JTC-FLAGERA Project GRANSPORT (M.I.K.). Support by
	the Netherlands National Computing Facilities foundation
	(NCF), with funding from the Netherlands Organisation for
	Scientific Research (NWO), is gratefully acknowledged.
	
	\appendix
	
    	\begin{center}
    	%\appendix
    	\chapter{\textbf{\large
    			Appendix
    	}}
    	
    \end{center}
    \section{Geometric interpretation of the model}
    \label{sec:geometry}
    \setcounter{equation}{0}
    \renewcommand{\theequation}{A.\arabic{equation}}
    
    The density of states of the Hamiltonian can be calculated via traces of Hamiltonian in some power, which can be expressed by the number of connected paths in a graph corresponding to the Hamiltonian. In the case of Cartesian product of graphs one can estimate the trace of a matrix power as:
    
    \begin{equation}
    \tr(H\square G)^m=p_m(H\square G)\sim\sum^{m}_{l=0} C^l_m p_{m-l}(G)p_l(H)
    \end{equation}
    where $p_n$ is the number of loops and index $n$ is the number of sites in this loop (length of the loop). This expression is exact, if each point in graphs $A$ or $H$ is indistinguishable. The full expression of $p_{m}$ is:
    
    \begin{equation}\label{eq:trace_paths}
    p_m(H\square G)=\sum_{x\in A, H}\sum^{m}_{l=0} C^l_m p_{m-l}(G_x)p_l(H_x)
    \end{equation}
    
    The block structure of a fractal graph can be represented by tensor product, which is closely related to the Cartesian product. The expressions \eqref{eq:fractal_sum}, \eqref{eq:fracta_sum_weighted} have tensor structures. From the geometric point of view, these formulas can be derived in the following. A point $p_0$ in $H\square G$ can be projected into $H$ or $G$, so $p_0$ has two coordinates. If we want to create a path between $p_0$ and another point $p_1$, we can project this path onto coordinates in $H$ or in $G$. By the structure of Cartesian product, we can always choose coordinates in $H$ or in $G$, and can combine closed paths and obtain Eq. (\ref{eq:trace_paths}).

    Let's consider the case, when the number of connections between two neighboring copies of graph $H$ is less than the number of vertices (i.e. the number of connections in the Cartesian product). Let's denote this matrix as $H*G$. In this case, a point $p_0$ also has two coordinates, however, we can not change make a new step in each of projections at arbitrary points. But we can make an estimation, saying that the number of closed paths in $G$ coordinate will be proportional to the number of paths in Eq. (\ref{eq:trace_paths}). With this approximation, we neglect details of the geometry and use only the number of connections between blocks. The coefficient of proportionality $\epsilon$ will be equal to fraction $n_c/n_v$, where $n_c$ is the number of connections and $n_v$ is the number of vertices in a graph $H$.
    
    \begin{gather}\label{eq:cart_estim}
    \tr(H*G)^m = p_m(H*G)\sim p_m(H) p_0(G)+
    \\+\sum^{m-1}_{l=0} \epsilon^{m-l} C^l_m p_{m-l}(G)p_l(H)\nonumber
    \end{gather}

    If a graph $G$ can be embedded into a graph $E$, it is obvious relation that $p_n(G)\le p_n(E)$. Because of the block structure, a fractal can be embedded (at least locally) into Cartesian product and we can apply this inequality. For fractals, roughly connections between blocks on different scales are described by matrix $A_0=a$. If we directly apply expression (\ref{eq:cart_estim}), it will correspond to $c\sim 1$ in Eq. (\ref{eq:fracta_sum_weighted}). Although this is also an estimation, but it doesn't include mixing of different scales on density of states (mixing of scales appears when one try to estimate traces from the Eq. (\ref{eq:fractal_construct_full}) due to noncommutativity). The model with $c\sim 1$ describes splitting of eigenvalues with the same order every time.
    
    In order to add an influence of each scale to another, we can say that, when we construct a path, every step on a larger scale is also a step in a smaller scales, but with some weight $\epsilon$. Then the trace of $A^m_2$: 
    
    \begin{equation}
    \tr(A^m_2)\sim p_m(a) n^m_v+\sum^{m-1}_{l=0} \epsilon^{m-l} C^l_m p_{m-l}(a)p_m(a)
    \end{equation}
    
    The formulas for greater iterations of fractal are cumbersome, but from the main text it is already clear this case corresponds to the Eq. (\ref{eq:fracta_sum_weighted}) and (\ref{eq:sum_spectrum}) with $c\sim a$.
    
    We can formulate the model of this article as follows. We build an effective model for the density of states of a fractal assuming that we only know  number of connections from one block to other.
    
    Of course, there can be other effective models with various weights on different scales. However, the model considered in this article clearly exploits scale symmetry of a system. If the detailed geometry does not have strict scale symmetry (for example, connections between blocks are in different places in every scale), then appropriate weighting of paths could be different, or the non-linearity could play stronger role.

\end{document}